\documentclass[aps,pre,showpacs,groupedaddress]{revtex4}

\usepackage{amsfonts,amssymb,amsmath,amsbsy}
\usepackage{epsfig}

\begin{document}

\title{Cavity approach to the Sourlas code system}

\author{Haiping Huang$^{1,2}$}
\affiliation{$^1$Key Laboratory of Frontiers in Theoretical Physics
and $^2$Kavli Institute for Theoretical Physics China, Institute of
Theoretical Physics, Chinese Academy of Sciences, Beijing 100190,
China}
\author{Haijun Zhou$^{1,2}$}
\affiliation{$^1$Key Laboratory of Frontiers in Theoretical Physics
and $^2$Kavli Institute for Theoretical Physics China, Institute of
Theoretical Physics, Chinese Academy of Sciences, Beijing 100190,
China}
\date{\today}

\begin{abstract}
The statistical physics properties of regular and irregular Sourlas
codes are investigated in this paper by the cavity method. At finite
temperatures,  the free energy density of these coding systems is
derived and compared with the result obtained by the replica method.
In the zero temperature limit, the Shannon's bound is recovered in
the case of infinite-body interactions while the code rate is still
finite. However, the decoding performance as obtained by the replica
theory has not considered the zero-temperature entropic effect. The
cavity approach is able to consider the ground-state entropy. It
leads to a set of evanescent cavity fields propagation equations
which further improve the decoding performance, as confirmed by our
numerical simulations on single instances. For the irregular Sourlas
code, we find that it takes the trade-off between good dynamical
property and high performance of decoding. In agreement with the
results found from the algorithmic point of view, the decoding
exhibits a first order phase transition as occurs in the regular
code system with three-body interactions. The cavity approach for
the Sourlas code system can be extended to consider first-step
replica-symmetry-breaking.
\end{abstract}

\pacs{02.70.-c, 89.90.+n, 89.70.-a, 05.50.+q}
 \maketitle

\section{Introduction}

Efficient and reliable transmission of information in noisy
environment plays a central role in modern information society.
Error-correcting codes, as efficient encoding/decoding mechanisms,
find widespread applications ranging from the satellite
communication to the storage of information on hard disks. In 1948,
Claude Shannon \cite{Shannon} proved that error-free transmission is
possible as long as the code rate $R$ (the ratio between the number
of bits in the original message and the number of bits in the
transmitted message) doesnot exceed the capacity of the channel
(Shannon's bound). More explicitly, for the binary symmetric channel
(BSC) where each transmitted bit is flipped independently with flip
 rate $p$, the Shannon bound is expressed as $R_{c}=1-H_{2}(p)$ where
$H_{2}(p)=-p\log_{2}p-(1-p)\log_{2}(1-p)$ is the binary entropy in
the information theory literature \cite{Elements}. This celebrated
channel encoding theorem forms the core of information theory.
However, it doesnot tell us how to construct an optimal code that
saturates Shannon's bound. In information science many efforts have
been devoted to construct  (near) optimal  codes  \cite{Galla}.

Based on insights gained from the study of disordered
systems \cite{spin_glass} the Sourlas code was proposed twenty years
ago, which relates error-correcting codes to spin glass
models \cite{Sourlas}. In the past decade, the statistical mechanics
analysis of Sourlas codes has been successfully generalized to other
types of error-correcting codes including low-density parity-check
(LDPC) codes, MacKay-Neal codes, Turbo codes, etc. Methods of
statistical physics, complementary to those used in information
theory, enable one to attain a more complete picture of decoding
process by analyzing global properties of the corresponding free
energy landscape. They also allow one to optimize the
performances of various codes by changing some construction parameters.

The procedure of constructing a Sourlas code is very simple. To
infer which bit is flipped by noise at the receiving end of
transmission, one has to introduce redundancy to the original
message at the sending end. As for the Sourlas code, the redundancy
is introduced by the Boolean sum of randomly selected message bits.
Through the transformation $\xi_{i}=(-1)^{x_{i}}$ where $x_{i}$ is
the Boolean bit and $\xi_{i}$ the Ising spin, the original bit
sequence $\{x_{i}\}$ can be regarded as an Ising spin configuration
$\{\xi_{i}\}$. In this way, the modulo $2$ addition is equivalent to
spin multiplication; and then the Sourlas code can be mapped to a
many-body spin glass problem \cite{Nishi_book}. In a general
scenario, the original message is an N-dimensional vector
$\boldsymbol{\xi}\in\{\pm1\}^{N}$, $M (>N)$ sets of interactions are
constructed by taking the product of randomly sampled $K$ bits from
the sequence of the original message, i.e.,
$J_{a}^{0}=\xi_{a_1}\cdots\xi_{a_K}$ $(a=1,\ldots, M)$. Then they
are fed into the noisy channel. At the destination, $M$ corrupted
interactions $J_{a}$, some of which being different from those at
the sending end, are received. The arising problem is how to infer
the original bits from the knowledge of channel outputs, statistical
properties of the channel and of the source. In the presence of weak
noise, searching for the ground state of the corresponding spin
glass model with given outputs $\{J_{a}\}$ will lead to successful
decoding. This decoding scheme is nothing but  maximum {\em a
posterior} probability (MAP) decoding. When the noise becomes
strong, the finite temperature decoding or marginal {\em posterior}
maximizer (MPM) scheme should be adopted since the ground state
would probably contain no information about the original message
\cite{Nishi_book,Rujan}.

The fully-connected Sourlas code has been studied in
Ref.~\cite{Sourlas}. It was shown that the Shannon's bound is
achieved in the limit $R\rightarrow 0$. Obviously, its practical
potential is greatly limited. The finite rate Sourlas code, of
greater practical significance, has been studied later on (see,
e.g., Refs.~\cite{Kaba01,Kaba02,Kaba03}). It turns out that at
finite coding rate $R$ the Shannon's bound for the channel capacity
can be attained at zero temperature at the limit of $K\rightarrow
\infty$ \cite{Kaba01,Kaba03}. However, the Shannon's bound couldnot
be achieved for finite $K$ despite its practical significance. All
the aforementioned investigations rely upon the replica method
developed initially for solving the Sherrington-Kirkpatrick model of
spin glass \cite{spin_glass,SK}. Moreover, they are restricted to
the replica symmetry (RS) assumption due to the emerging more
complicated saddle point analysis of replica symmetry breaking.
Nevertheless, recent developments in the study of LDPC codes
\cite{Migli} showed that the one-step replica symmetry breaking
(1RSB) type algorithm is able to shift the dynamical phase
transition \cite{Franz02} to a higher value as compared with RS type
algorithms. Similar results were obtained on the finite connectivity
Sourlas code system from the dynamic point of view \cite{Kap,Hat}.
In this work we study the equilibrium properties of the finite
connectivity Sourlas code system by using the cavity method of
statistical physics \cite{Mezard_book,Cavity,SAT}.

The cavity method has its own advantages over the replica method.
The latter is based on a saddle point analysis of $n$-dimensional
integral in the limit $n\rightarrow 0$. This analytic continuation
in the number of replicas hasn't been confirmed to hold generally,
neither has the validity of the exchange of the order of two limits
($N\rightarrow\infty$ and $n\rightarrow 0$ ). On the other hand, the
cavity method adopts a direct probabilistic analysis, which makes it
applicable to single problem instances.  In this paper, it is
expected that the cavity method reproduces results obtained by
replica theory. Within the cavity framework, the entropic
contribution in the zero temperature limit can be taken into account
by means of first order corrections in temperature $T$, which has
led to interesting insights on the ground state solution space
properties of several disordered systems such as the random vertex
cover problem and the random  matching problem \cite{Entropy}.
Following the same strategy, we derive the evanescent cavity fields
propagation (ECFP) equation for decoding Sourlas codes, and find it
outperforms the traditional case where only the hard field or
energetic contribution is considered.

The rest of this paper is organized as follows. The model is
introduced in Sec.~\ref{sec_model}. In Sec.~\ref{sec_CM}, iterative
equations for finite temperature decoding and zero temperature
decoding are rederived respectively using the cavity method. Taking
into account the entropic contribution, we also propose the ECFP
equation. In Sec.~\ref{sec_result}, regular (with a single $K$
value) and irregular (with several values of $K$) Sourlas codes are
discussed.  In this section, it is also observed that the ECFP
procedure is able to improve the decoding performance by a
significant amount. We conclude this paper in Sec.~\ref{sec_Con} and
make further discussions there.

\begin{center}
\begin{figure}
(a)\epsfig{file=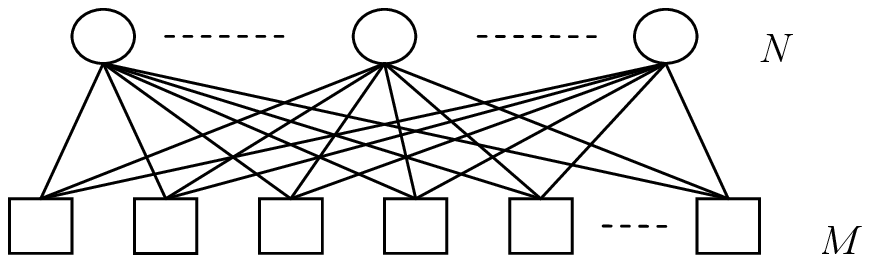,bb=101 633 475 767,width=7cm} \hskip 1cm
(b)\epsfig{file=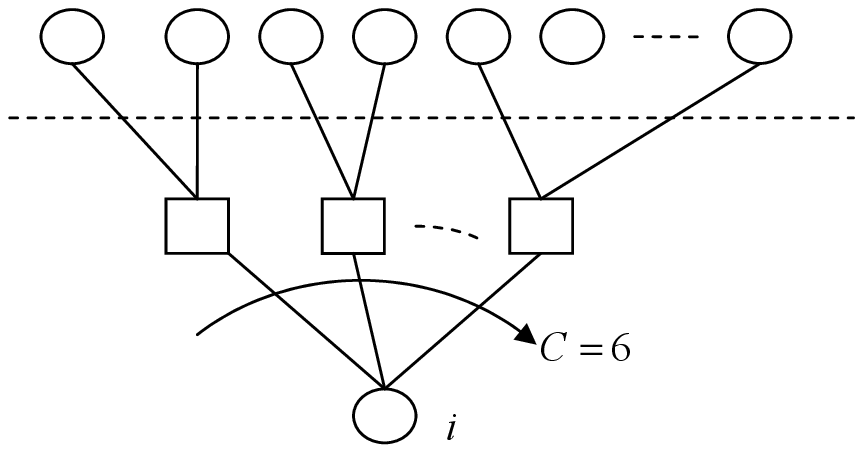,bb=119 619 429 736,width=7cm} \vskip 1cm
(c)\epsfig{file=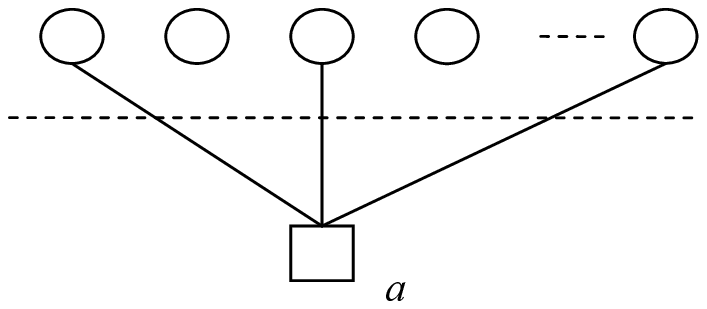,bb=127 612 432 736,width=7cm} \hskip
.5cm \caption{ Factor graph representation of a random construction
of Sourlas codes (a) and the cavity method (b,c). (a) There are
totally $N$ bits (circles) and $M$ parity checks (squares) in the
factor graph. Each bit (variable node) is connected to exactly six
parity checks (function nodes), and each parity check involves three
bits. (b) A single new bit $i$ together with six parity checks is
added to the original system denoted by the part above the dashed
line. (c) A new function node $a$ connected to three randomly
selected bits is added to the original system.} \label{Cavity}
\end{figure}
\end{center}

\section{Model}
\label{sec_model}

Hereafter, we adopt the Ising spin representation of the Boolean
numbers. In the Sourlas code scenario, the original binary message
$\boldsymbol{\xi}\in\{\pm1\}^{N}$ of length $N$ is encoded into a
transmitted binary message $\boldsymbol{J}^0= \{J_1^0, J_2^0,
  \ldots, J_M^0\}$ of length $M$, with the $a$-th bit $J_a^0$
being the product of a subset $\partial a$ of the original message
bits, $J_a^0 = \prod_{i\in \partial a}\xi_{i}$ (see
Fig.~\ref{Cavity}(a) for a pictorial description, in which a parity
check $a$ is represented by a square and a message bit is
represented by a circle). If each parity check involves $K$ bits and
each bit is constrained by $C$ parity checks, then the coding rate
is $R \equiv \frac{N}{M}=\frac{K}{C}$. The Hammitonian of the system
reads
\begin{equation}\label{Hami}
    \mathcal {H}=-\sum_{a=1}^{M}J_{a}\prod_{i\in\partial a}\sigma_{i}
\end{equation}
where $\{\sigma_{i}\}$ are referred to as dynamical spin variables
for decoding and $\{J_{a}\}^{M}$ is the received message. Due to the
noise in the transmission channel, the received message may not be identical
to the transmitted one $\{J_{a}^{0}\}$. We assume memoryless
binary symmetric channel, i.e.,
\begin{equation}\label{BSC}
    P(J_{a}|J_{a}^{0})=p\delta(J_{a}+J_{a}^{0})+(1-p)\delta(J_{a}-J_{a}^{0})
\end{equation}
where $p$ is the flip rate.

Introducing an inverse temperature $\beta$ as a control parameter,
the spin configuration $\boldsymbol{\sigma}$ is sampled with
probability
\begin{equation}
  P(\boldsymbol{\sigma}|\boldsymbol{J})=\frac{\exp\bigl(-\beta \mathcal{H} (\boldsymbol{\sigma})\bigr)}
    {Z}.
\end{equation}
where $Z$ is the partition function. On the other hand, the
Hamiltonian (\ref{Hami}) is invariant under the gauge transformation
$\sigma_{i}\rightarrow\sigma_{i}\xi_{i}$, $J_{a}\rightarrow
J_{a}\prod_{i\in\partial a}\xi_{i}$. Therefore, any general message
can be mapped onto a ferromagnetic configuration $\{\xi_{i}=+1\}$.
Under this transformation, Eq.~(\ref{BSC}) can be re-written as
\begin{equation}\label{BSC2}
    P(J_{a})=p\delta(J_{a}+1)+(1-p)\delta(J_{a}-1)
\end{equation}
In this sense, the Sourlas code is actually a multi-spin
ferromagnetically biased $\pm J$ spin glass model.

The aim of the statistical inference problem is to estimate the
marginal posterior $P(\sigma_{i}|\boldsymbol{J})$. We adopt the MPM
estimator
$\hat{\xi_{i}}={\rm sgn}\bigl(P(\sigma_{i}=1|\boldsymbol{J})
-P(\sigma_{i}=-1|\boldsymbol{J})\bigr)={\rm sgn}\bigl<\sigma_{i}\bigr>_{\beta}$.
To measure the performance of decoding, one usually defines the
overlap between the estimated bits $\{\hat{\xi_{i}}\}$ and the
original message $\{\xi_{i}\}$ as
\begin{equation}\label{overlap}
    m(\beta,p)=\frac{1}{N}\sum_{i=1}^{N}\xi_{i}\hat{\xi}_{i}=\frac{1}{N}\sum_{i=1}^{N}\xi_{i}
{\rm sgn}\bigl<\sigma_{i}\bigr>_{\beta}
\end{equation}
where ${\rm sgn}(x)=x/|x|$ for $x\neq 0$. Evaluating
$\bigl<\sigma_{i}\bigr>_{\beta}$ directly is computationally
expensive, however, it can be well approximated using the cavity
method presented in the next section. If we focus on typical value
of the decoding overlap, Eq.~(\ref{overlap}) should be averaged over
the quenched disorder, i.e.,
\begin{equation}
    m(\beta,p)=\frac{1}{N}\left<\sum_{i=1}^{N}\xi_{i}
    \biggl<{\rm sgn}\bigl<\sigma_{i}\bigr>_{\beta}\biggr>_{\mathfrak{C},
P(\boldsymbol{J}|\boldsymbol{\xi})}\right>_{\boldsymbol{\xi}}
\end{equation}
where $\mathfrak{C}$ represents the average over random
constructions of codes with fixed bit's degree $C$. The other two
types of quenched disorder come from the corruption process
($P(\boldsymbol{J}|\boldsymbol{\xi})$) and the distribution of the
original message bits $P(\boldsymbol{\xi})$. For simplicity, we
concentrate on typical properties of the system with unbiased
original message and memoryless binary symmetric channel. In the
long message limit ($N\rightarrow \infty$), it is believed that the
macroscopic observables for a given instance are independent of the
particular realization of the disorder \cite{spin_glass,Kaba03}.

\section{Cavity Method}
\label{sec_CM}

Using the replica method, one is forced to work directly
with the disorder average from the start, whereas the cavity method
admits of taking the average over the quenched disorder after the
computation. In this section, we derive the free energy at finite
temperature as well as zero temperature for the finite connectivity
Sourlas code system using the cavity method, and then extend the
result to the irregular Sourlas code case. Within the cavity
framework, the entropic contribution is considered in the zero
temperature limit and the ECFP equation is proposed as well.

\subsection{Finite temperature decoding}
\label{subsec:FTD}

Because of the random construction of Sourlas codes, it is
reasonable to assume that the correlation between randomly sampled
bits vanishes in the long message limit. We assume all the
calculations below are within the RS ansatz (single-state cavity
method). The results are straightforward to be generalized to 1RSB
case.

As shown in Fig.~\ref{Cavity}(b), if we add one variable node to the
original system, $C$ function nodes should be added simultaneously.
Then the partition function for the enlarged system is:
\begin{widetext}
\begin{equation}\label{Z_Node}
    \begin{split}
    Z^{new}&=\sum_{\sigma_{i}}\sum_{\vec{\sigma}}\exp\bigl({\sum_{a=1}^{M}\beta J_{a}\prod_{k\in\partial a}\sigma_{k}+
    \beta\sum_{b=1}^{C}J_{b}\sigma_{i}\prod_{j\in\partial b\backslash i}\sigma_{j}}\bigr)\\
    &=Z^{old}\sum_{\sigma_{i}}\prod_{b}\sum_{\{\sigma_{j}\}:j\in\partial b\backslash i}\prod_{j\in \partial b\backslash i}
    \biggl[\frac{e^{\beta h_{j\rightarrow b}\sigma_{j}}}{2\cosh\beta h_{j\rightarrow
    b}}\biggr]\cdot e^{\beta J_{b}\sigma_{i}\prod_{j\in \partial b\backslash
    i}\sigma_{j}}\\
    &=Z^{old}\biggl[\prod_{b}\bigl[\cosh\beta J_{b}\bigl(1+\tanh\beta J_{b}\prod_{j\in\partial b\backslash i}\tanh\beta h_{j\rightarrow
    b}\bigr)\bigr]+\prod_{b}\bigl[\cosh\beta J_{b}\bigl(1-\tanh\beta
J_{b}\prod_{j\in\partial b\backslash i}\tanh\beta h_{j\rightarrow
b}\bigr)\bigr]\biggr]
    \end{split}
\end{equation}
\end{widetext}
where $\sigma_{i}$ is the newly added spin,
$Z^{old}=\sum_{\vec{\sigma}}\exp({\sum_{a=1}^{M}\beta J_{a}\prod_{i\in
\partial a}\sigma_{i}})$ is the partition function of the old system,
 $h_{j\rightarrow b}$ is the cavity field of variable node $j$
when function node $b$ is removed from the graph, $j\in\partial
b\backslash i$ denotes the set of bits involved in function node $b$
but $i$ is excluded from this set. To derive the second equality in
Eq.~(\ref{Z_Node}), we have made use of the absence of strong
correlation between randomly chosen spins, since for one random
construction of Sourlas codes depicted in Fig.~\ref{Cavity}(a), the
typical loop size in the corresponding factor graph is of order
$\log N$ which diverges in $N\rightarrow \infty$. In this sense, the
joint probability of a few randomly selected spins
$P(\boldsymbol{\sigma}_{\partial a})$ is factorized as
$P(\boldsymbol{\sigma}_{\partial a})\approx \prod_{i\in\partial
a}P(\sigma_{i})$ where we write single node belief $P(\sigma_{i})$
as $P(\sigma_{i})=\frac{e^{\beta h_{i}\sigma_{i}}}{2\cosh \beta
h_{i}}$ in terms of the local field $h_{i}$ acting on the spin
$\sigma_{i}$.

Upon defining the magnetization $m_{i\rightarrow b}\equiv \tanh\beta
h_{i\rightarrow b}$ and the conjugate magnetization
$\hat{m}_{b\rightarrow i}\equiv\tanh \beta u_{b\rightarrow
i}\equiv\tanh\beta J_{b}\prod_{j\in\partial b\backslash i}\tanh
\beta h_{j\rightarrow b}$ where $u_{b\rightarrow i}$ is termed the
cavity bias, one gets the free energy shift due to one variable node
addition:
\begin{widetext}
\begin{equation}\label{deltaFi}
    -\beta\Delta F_{i}=\log\frac{Z^{new}}{Z^{old}}=\log\biggl[\prod_{b\in\partial i}\bigl[\cosh\beta J_{b}(1+\hat{m}_{b\rightarrow i})
    \bigr]+\prod_{b\in\partial i}\bigl[\cosh\beta J_{b}(1-\hat{m}_{b\rightarrow i})\bigr]\biggr]
\end{equation}

As the second step, one function node addition is performed (c.f.
Fig.~\ref{Cavity}(c)). Likewise, the new partition function reads
\begin{equation}\label{Z_Func}
    \begin{split}
    Z^{new}&=\sum_{\vec{\sigma}}\exp\bigl({\beta\sum_{a=1}^{M}J_{a}\prod_{k\in \partial
    a}\sigma_{k}+\beta J_{a}\prod_{i\in\partial a}\sigma_{i}}\bigr)\\
    &=\sum_{\vec{\sigma}}e^{\beta\sum_{a=1}^{M}J_{a}\prod_{k\in \partial
    a}\sigma_{k}}\sum_{\vec{\sigma}}\frac{e^{\beta\sum_{a=1}^{M}J_{a}\prod_{k\in \partial
    a}\sigma_{k}}}{\sum_{\vec{\sigma}}e^{\beta\sum_{a=1}^{M}J_{a}\prod_{k\in \partial
    a}\sigma_{k}}}e^{\beta J_{a}\prod_{i\in \partial a}\sigma_{i}}\\
    &=Z^{old}\sum_{\vec{\sigma}}P(\vec{\sigma})e^{\beta J_{a}\prod_{i\in \partial
    a}\sigma_{i}}\\
    &=Z^{old}\sum_{\{\sigma_{i}\}:i\in\partial a}\prod_{i\in\partial
    a}\biggl[\frac{e^{\beta h_{i\rightarrow a}\sigma_{i}}}{2
    \cosh\beta h_{i\rightarrow a}}\biggr]e^{\beta J_{a}\prod_{i\in \partial
    a}\sigma_{i}}\\
    &=Z^{old}\cdot\cosh\beta J_{a}\bigl(1+\tanh\beta J_{a}\prod_{i\in\partial a}m_{i\rightarrow a}\bigr)
    \end{split}
\end{equation}
\end{widetext}
The corresponding free energy shift is $-\beta\Delta
F_{a}=\log\biggl[\cosh\beta J_{a}\bigl(1+\tanh\beta
J_{a}\prod_{i\in\partial a}m_{i\rightarrow a}\bigr)\biggr]$.
Finally the total free energy density is given by \cite{Cavity}
\begin{equation}\label{Total_F}
    \begin{split}
    f&=\frac{1}{N}\sum_{i}\Delta
    F_{i}-\frac{1}{N}\sum_{a}(|\partial a|-1)\Delta F_{a}\\
    &=\bigl<\Delta F_{i}\bigr>_{pop}-\frac{K-1}{K}C\bigl<\Delta F_{a}\bigr>_{pop}
    \end{split}
\end{equation}
where $\bigl<\cdots\bigr>_{pop}$ means the average over populations
of $\{m_{i\rightarrow a},\hat{m}_{a\rightarrow i}\}$ when the
population dynamics recipe \cite{Cavity} is adopted. The second term
in the final expression of Eq.~(\ref{Total_F}) can be understood as
follows: When one variable node is added, the number of
over-generated function nodes is $\frac{K-1}{K}C$ on average; the
contribution of these nodes should be eliminated from the total free
energy. Following the same line mentioned above, one can write
$m_{i\rightarrow a}$ as a function of $\{\hat{m}_{b\rightarrow
i}\}_{b\in\partial i\backslash a}$, then obtain a closed set of
equations in the form of distribution:
\begin{widetext}
\begin{subequations}\label{BP}
\begin{align}
  P(m_{i\rightarrow a})=\int\left[\prod_{b\in\partial i\backslash a}Q(\hat{m}_{b\rightarrow i})d\hat{m}_{b\rightarrow
  i}\right]
  \delta\left(m_{i\rightarrow a}-\frac{\prod_{b\in\partial i\backslash a}(1+\hat{m}_{b\rightarrow i})-
  \prod_{b\in\partial i\backslash a}(1-\hat{m}_{b\rightarrow i})}{\prod_{b\in\partial i\backslash a}(1+\hat{m}_{b\rightarrow
  i})+
  \prod_{b\in\partial i\backslash a}(1-\hat{m}_{b\rightarrow i})}\right)\\
Q(\hat{m}_{b\rightarrow i})=\int\left[\prod_{j\in\partial
b\backslash i}P(m_{j\rightarrow b})dm_{j\rightarrow
b}\right]\delta\left(\hat{m}_{b\rightarrow i}-\tanh\beta
J_{b}\prod_{j\in\partial b\backslash i}m_{j\rightarrow b}\right)
  \end{align}
\end{subequations}
\end{widetext}
Eq.~(\ref{BP}) is nothing but the belief propagation equation when
applied to a single instance (one particular realization of Sourlas
codes) \cite{Kaba02}. Population dynamics recipe is applied to solve
the recursive equations above. When the iteration reaches a steady
state, the free energy can be computed and the marginal posterior
can be well approximated by
$P(\sigma_{i})=\frac{1+m_{i}\sigma_{i}}{2}$ for the sparse random
graph. According to Eq.~(\ref{overlap}), the performance of decoding
is evaluated via $m=\frac{1}{N}\sum_{i}\xi_{i}\hat{\xi}_{i}=\int
dm_{i}P(m_{i}){\rm sgn}(m_{i})$,
where the gauge transformation has been performed and the
magnetization $m_{i}$ obeys the distribution
\begin{widetext}
\begin{equation}\label{Prob_mi}
    P(m_{i})=\int\left[\prod_{b\in\partial i}Q(\hat{m}_{b\rightarrow i})d\hat{m}_{b\rightarrow
    i}\right]
  \delta\left(m_{i}-\frac{\prod_{b\in\partial i}(1+\hat{m}_{b\rightarrow i})-
  \prod_{b\in\partial i}(1-\hat{m}_{b\rightarrow i})}{\prod_{b\in\partial i}(1+\hat{m}_{b\rightarrow
  i})+
  \prod_{b\in\partial i}(1-\hat{m}_{b\rightarrow i})}\right)
\end{equation}
\end{widetext}

\subsection{Zero temperature decoding}
\label{subsec:ZTD}

The finite temperature decoding is facilitated through
Eq.~(\ref{BP}). However, searching for the ground state of the
system requires performing zero temperature decoding, and the
equations derived above can be further simplified. Taking the limit
$\beta\rightarrow \infty$, one obtains the recursive equations for
cavity fields and biases:
\begin{widetext}
\begin{subequations}\label{ZTBP}
\begin{align}
  P(h_{i\rightarrow a})=\int\left[\prod_{b\in\partial i\backslash a}du_{b\rightarrow i}
  Q(u_{b\rightarrow i})\right]\delta\left(h_{i\rightarrow a}-\sum_{b\in\partial i\backslash a}u_{b\rightarrow i}\right)\label{ZTBP01} \\
  Q(u_{b\rightarrow i})=\int\left[\prod_{j\in\partial b\backslash i}dh_{j\rightarrow b}
  P(h_{j\rightarrow b})\right]\delta\left(u_{b\rightarrow i}-{\rm sgn}\bigl(J_{b}\prod_{j\in\partial b\backslash i}h_{j\rightarrow
  b}\bigr)\right)\label{ZTBP02}
  \end{align}
\end{subequations}
and the free energy shifts
\begin{subequations}
\begin{align}
  -\Delta F_{i}=C-\sum_{b\in\partial i}\left|u_{b\rightarrow i}\right|+\left|\sum_{b\in\partial i}u_{b\rightarrow i}\right| \\
  -\Delta F_{a}=1-2\Theta\left(-J_{a}\prod_{i\in\partial a}h_{i\rightarrow
  a}\right)\label{ZTF02}
  \end{align}
\end{subequations}
\end{widetext}
where $\Theta(x)$ is a step function taking values $\Theta(x)=0$ for
$x\leq 0$, $\Theta(x)=1$ for $x>0$. In Eq.~(\ref{ZTBP02}), we take
the convention ${\rm sgn}(0)=0$. Similarly, the overlap in the
zero-temperature limit reads $m=\int dhP(h){\rm sgn}(h)$ where the
field is subject to the distribution
$P(h)=\int\left[\prod_{b\in\partial i}Q(u_{b\rightarrow
i})du_{b\rightarrow i}\right]\delta\left(h-\sum_{b\in\partial
i}u_{b\rightarrow i}\right)$ where $Q(u_{b\rightarrow i})$ is the
distribution of cavity biases according to Eq.~(\ref{ZTBP02}).

\subsection{Evanescent cavity fields propagation}
\label{subsec:ECFP}

In Sec.~\ref{subsec:ZTD}, only the hard field or energetic
contribution is considered. We expect that the neglected entropic
contribution will provide useful information for improving the
decoding performance. To derive the ECFP equation, we rewrite
Eq.~(\ref{BP}) in terms of cavity fields:
\begin{equation}\label{WP}
    \eta_{i\rightarrow a}\equiv 2h_{i\rightarrow a}=\sum_{b\in\partial i\backslash
    a}\frac{1}{\beta}\log\left[\frac{1+\hat{m}_{b\rightarrow i}}{1-\hat{m}_{b\rightarrow i}}\right]
\end{equation}
When we consider only the energetic contribution in the zero
temperature limit, the resulting closed set of equations
Eq.~(\ref{ZTBP}) are called warning propagation (WP) \cite{Entropy}.
The limit $\beta\rightarrow\infty$ selects the ground state of the
system under consideration, therefore WP also corresponds to the MAP
estimator. However, as $T$ goes to zero, the local field $h_{i}$
vanishes linearly in $T$, consequently contributes to the
corresponding local magnetization \cite{SAT}. That is to say, even
if the local field takes value of zero, the non-vanishing evanescent
part, defined as the coefficient of first order correction of cavity
field with respect to $T$, still results in a finite magnetization.
Therefore, these evanescent fields are expected to provide useful
information for improving the decoding performance. Expanding the
cavity field $h_{i\rightarrow a}$ up to the first order in $T$,
i.e.,
\begin{equation}\label{Expfiled}
    \eta_{i\rightarrow a}=2I_{i\rightarrow a}+\frac{r_{i\rightarrow a}}{
    \beta}
\end{equation}
where $I_{i\rightarrow a}$ is an integer corresponding to the
energetic contribution and $r_{i\rightarrow a}$ a finite real value
corresponding to the entropic contribution, then substituting
Eq.~(\ref{Expfiled}) into Eq.~(\ref{WP}), one readily gets ECFP
equations:
\begin{widetext}
\begin{subequations}\label{ECFPeq}
\begin{align}
  I_{i\rightarrow a}&=\sum_{b\in\partial i\backslash a}{\rm sgn}(J_{b}\prod_{j\in\partial
b\backslash i}I_{j\rightarrow b})\\
  \begin{split}
  r_{i\rightarrow a}&=\sum_{b\in\partial i\backslash
  a}\mathbb{I}\left(I_{j\rightarrow b}=0\;\forall j\in\partial b\backslash i\right)\log\left[\frac{\prod_{j\in\partial b\backslash i}
  (e^{r_{j\rightarrow b}}+1)+J_{b}\prod_{j\in\partial b\backslash i}
  (e^{r_{j\rightarrow b}}-1)}{\prod_{j\in\partial b\backslash i}
  (e^{r_{j\rightarrow b}}+1)-J_{b}\prod_{j\in\partial b\backslash i}
  (e^{r_{j\rightarrow b}}-1)}\right]\\
&+\mathbb{I}\left(\text{at least one}\;I_{j\rightarrow b}=0,\text{at
most}\;(K-2)\;I_{j\rightarrow b}=0\;\forall j\in\partial b\backslash
i
\right)\\
&\cdot\log\left[\frac{\prod_{j\in\partial b\backslash i}^{'}
  (1+e^{-|r_{j\rightarrow b}|})+\tilde{J}_{b}\prod_{j\in\partial b\backslash i}^{'}
  (1-e^{-|r_{j\rightarrow b}|})}{\prod_{j\in\partial
b\backslash i}^{'}
  (1+e^{-|r_{j\rightarrow b}|})-\tilde{J}_{b}\prod_{j\in\partial b\backslash i}^{'}
  (1-e^{-|r_{j\rightarrow b}|})}\right]\\
&-\mathbb{I}\left(I_{j\rightarrow b}\neq0\;\forall j\in\partial
b\backslash i\right){\rm sgn}(J_{b}\prod_{j\in\partial b\backslash
i}I_{j\rightarrow b})\log(1+R_{b\rightarrow i})\label{ECFPeqr}
\end{split}
  \end{align}
\end{subequations}
\end{widetext}
where $\prod_{j\in\partial b\backslash i}^{'}=\prod_{\substack{j\in\partial b\backslash i\\
\{I_{j\rightarrow b}=0\}}}$,
$\tilde{J}_{b}=J_{b}{\rm sgn}\left(\prod_{\substack{k\in\partial b\backslash i\\
\{I_{k\rightarrow b}\neq0\}}}I_{k\rightarrow b}\right){\rm
sgn}\left(\prod_{j\in\partial b\backslash i}^{'} r_{j\rightarrow
b}\right)$, $\mathbb{I}(\cdot)$ is the indicator function of an
event and $R_{b\rightarrow i}=\sum_{j\in\partial b\backslash
i}\tilde{R}_{j\rightarrow b}$ where $\tilde{R}_{j\rightarrow
b}=\exp\left[-{\rm sgn}(I_{j\rightarrow b})r_{j\rightarrow
b}\right]$ if $|I_{j\rightarrow b}|=1$, and $0$ otherwise. In the
summation of Eq.~(\ref{ECFPeqr}), the first term corresponds to the
case where $I_{j\rightarrow b}=0$ for all $j\in\partial b\backslash
i$, the second term the case where at least one $I_{j\rightarrow
b}=0$, at most $(K-2)$ $I_{j\rightarrow b}=0$ and the last term the
case where $I_{j\rightarrow b}\neq0$ for all $j\in\partial
b\backslash i$. Then the decoding can be easily performed via
$m=\int dI_{i} dr_{i} P(I_{i}) Q(r_{i})\left({\rm
sgn}(I_{i})+\mathbb{I}(I_{i}=0){\rm sgn}(r_{i})\right)$ where $P,Q$
represent the distributions for the hard fields $\{I_{i}\}$ and
evanescent fields $\{r_{i}\}$ respectively when the population
dynamics technique is used to solve the ECFP equations. Actually, in
the zero temperature limit, the estimated message bit
$\hat{\xi_{i}}={\rm sgn}(m_{i})={\rm sgn}(\tanh\beta h_{i})={\rm
sgn}(I_{i})$ if $I_{i}\neq0$ and ${\rm sgn}(r_{i})$ otherwise.

\subsection{Decoding irregular Sourlas codes}
\label{subsec:DISC} All the aforementioned computations are limited
to the regular case where the check's degree $K$ takes a single
value. It is worthwhile to study the irregular case. The irregular
Sourlas code is defined as the code with various values of $K$. We
assume the check's degree $K$ follows a distribution with two
delta-peaks
\begin{equation}\label{irregular_K}
    P(K)=\gamma\delta(K-2)+(1-\gamma)\delta(K-3)
\end{equation}
We adopt this form of distribution for two reasons. One is the
Sourlas code has perfect dynamical properties for $K=2$ and high
decoding performance for $K=3$. The other is the result can be
compared with that obtained for the cascading Sourlas code
\cite{Hat,KS}. The formula for the total free energy density of the
combined system is given by
\begin{widetext}
\begin{equation}\label{Total_F_Irre}
    \begin{split}
    f&=\frac{1}{N}\sum_{i}\Delta
    F_{i}-\frac{1}{N}\sum_{a}(|\partial a|-1)\Delta F_{a}\\
    &=\left<\Delta F_{i}\right>_{pop}-\frac{M}{N}\sum_{K}P(K)(K-1)\left<\Delta F_{a}\right>_{pop}\\
    &=\left<\Delta F_{i}\right>_{pop}-\frac{C}{\overline{K}}\left[\gamma\left<\Delta F_{K=2}\right>_{pop}+2(1-\gamma)\left<\Delta
    F_{K=3}\right>_{pop}\right]
    \end{split}
\end{equation}
where $\overline{K}=3-\gamma$ and the code rate
$R=\frac{3-\gamma}{C}$. The recursive equations are of the form
\begin{subequations}\label{BP_Irre}
\begin{align}
  P(m_{j})=\int\left[\prod_{b=1}^{C-1}d\hat{m}_{b}Q(\hat{m}_{b})\right]\delta\left(m_{j}-\frac{\prod_{b=1}^{C-1}(1+\hat{m}_{b})
  -\prod_{b=1}^{C-1}(1-\hat{m}_{b})}{\prod_{b=1}^{C-1}(1+\hat{m}_{b})
  +\prod_{b=1}^{C-1}(1-\hat{m}_{b})}\right)\\
Q(\hat{m}_{b})=\sum_{K}\frac{P(K)K}{\overline{K}}\int\left[\prod_{j=1}^{K-1}P(m_{j})dm_{j}\right]\delta\left(\hat{m}_{b}-\tanh\beta
J_{b}\prod_{j=1}^{K-1}m_{j}\right)\label{BP_Irre02}
  \end{align}
\end{subequations}
\end{widetext}
Eq.~\ref{BP_Irre02} can be understood as follows: since $\gamma$
represents the fraction of function nodes with $2$-spin interaction,
for one randomly chosen bit, it is connected to a parity check
involving two bits with probability $P_{2}=\frac{2\gamma}{3-\gamma}$
and to that involving three bits with probability
$P_{3}=\frac{3(1-\gamma)}{3-\gamma}$. Obviously, $P_{2}+P_{3}=1$.
The formula for zero temperature decoding of irregular codes can be
derived similarly. In the next section, we will discuss the
performance of decoding for regular and irregular codes
respectively.

\begin{center}
\begin{figure}
(a)\epsfig{file=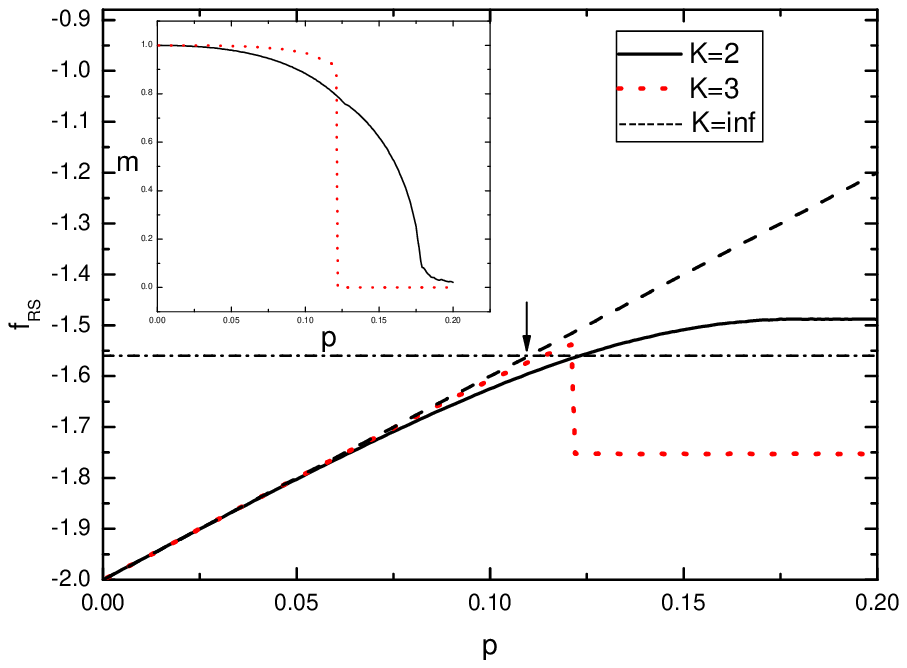,bb=12 12 282 215,width=8.0cm}\hskip
1cm (b)\epsfig{file=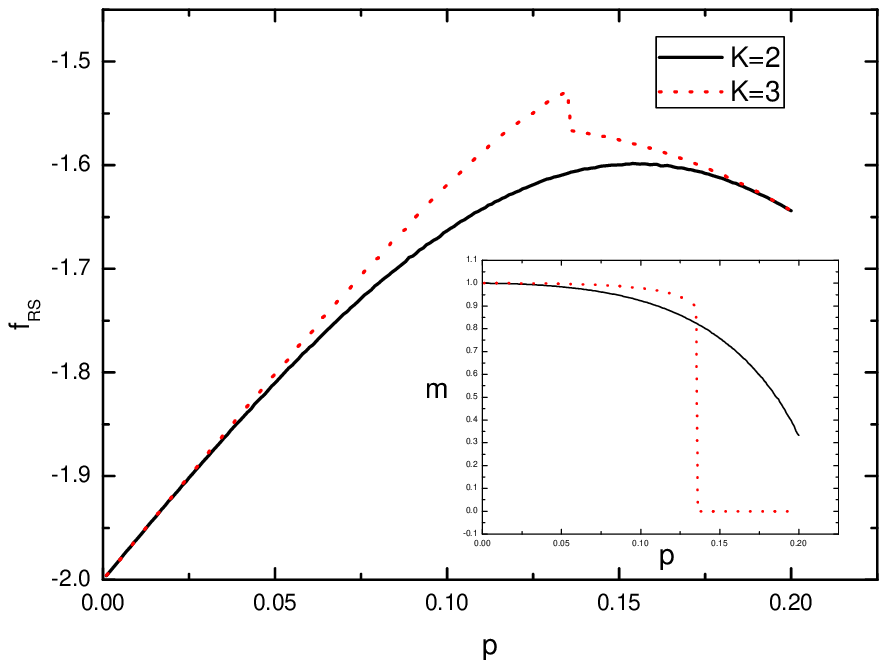,bb=13 13 278
218,width=8.0cm}\vskip .5cm
 \caption{(Color online) The decoding performance for regular Sourlas codes with
$R=0.5$. The calculated mean values are shown and the corresponding
variances are smaller than the symbol size. (a)The replica symmetry
free energy density versus flip rate when zero temperature decoding
is performed. The solid line corresponds to $K=2$ case while the
dotted line $K=3$ case. The dashed line represents the case of
$K\rightarrow\infty$, and the dashed-dotted line corresponds to the
1RSB frozen spins solution. The arrow indicates the critical noise
level where the Shannon's bound is achieved. (b)The replica symmetry
free energy density versus flip rate when finite temperature
decoding is performed. The decoding temperature is chosen to be
Nishimori temperature $\beta_{p}=\frac{1}{2}\log\frac{1-p}{p}$.
Insets: the overlap versus flip rate for zero temperature and finite
temperature decoding respectively.} \label{Regular}
\end{figure}
\end{center}

\begin{center}
\begin{figure}
\epsfig{file=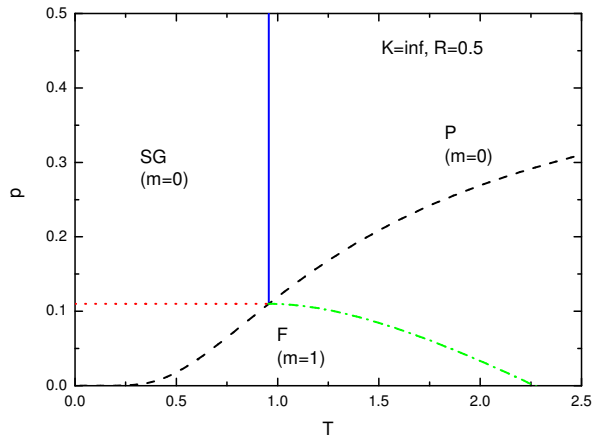,bb=13 13 278 215,width=8.0cm}
\caption{(Color online) The phase diagram for regular Sourlas codes
with $K\rightarrow \infty $ keeping $R=0.5$. The dashed line
indicates the Nishimori line, the dotted line the boundary between
spin glass (SG) phase and ferromagnetic (F) phase, the dashed-dotted
line the boundary between F and paramagnetic (P) phase and the solid
line the boundary between SG and P.} \label{PDK}
\end{figure}
\end{center}

\begin{center}
\begin{figure}
\epsfig{file=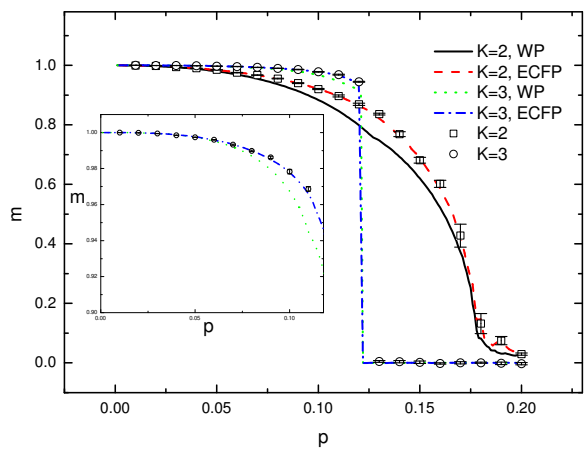,bb=14 15 273 214,width=8.0cm} \caption{(Color
online) The decoding overlap $m$ as a function of flip rate $p$ for
regular Sourlas codes with $R=0.5$. The solid or dashed line
corresponds to zero temperature decoding for the $K=2$ case while
the dotted or dashed dotted line the $K=3$ case. The solid (black)
or dotted (green) one represents results obtained by the
conventional warning propagation (WP) while the dashed (red) or
dashed dotted (blue) one evanescent cavity fields propagation
(ECFP). The cutoff takes the value $4.0$. Numerical simulations on
on a single graph by ECFP are consistent with the mean field
results. The size of the graph is $N=10000$. The decoding result on
single graph is averaged over ten individual simulations for each
flip rate $p$. The error bars indicate the standard deviations.
Inset: a detailed view of the significant improvement using ECFP for
the $K=3$ case.} \label{ECFP}
\end{figure}
\end{center}

\begin{center}
\begin{figure}
\epsfig{file=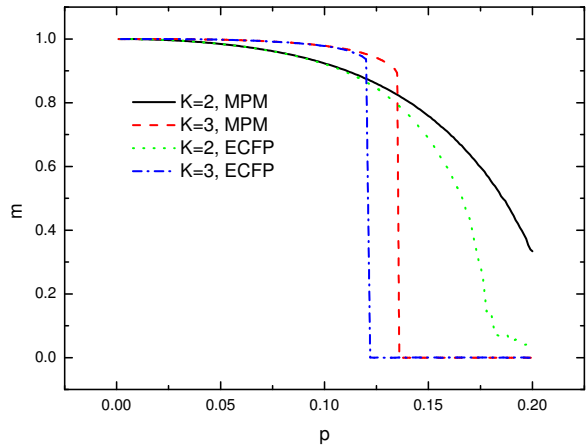,bb=14 15 272 211,width=8.0cm}
\caption{(Color online) The decoding overlap $m$ as a function of
flip rate $p$ for regular Sourlas codes with $R=0.5$. The calculated
mean values are shown and the corresponding variances are smaller
than the symbol size. The solid and dashed line correspond to finite
temperature decoding (MPM) for the $K=2$ case and $K=3$ case
respectively, while the dotted and dashed dotted line the ECFP
decoding for which the cutoff takes the value $4.0$.}
\label{MPMvsECFP}
\end{figure}
\end{center}


\section{Results and Discussions}
\label{sec_result}
\subsection{Regular Sourlas codes}\label{sec_RSC}
Properties of regular Sourlas codes have been studied using replica
theory \cite{Kaba01,Kaba03}. In this section, we reproduce results
obtained on regular Sourlas codes on the basis of the cavity method.

For regular Sourlas codes, we consider the case of $K=2$ and $K=3$
with the same code rate $R=0.5$. In particular, the other cases
($K>3$) show the same qualitative behavior as the $K=3$ case. It is
worthwhile to mention that Eq.~(\ref{BP}) yields a paramagnetic
solution, i.e., $P(m_{i\rightarrow a})=\delta(m_{i\rightarrow a}),
Q(\hat{m}_{a\rightarrow i})=\delta(\hat{m}_{a\rightarrow i})$.
Following Eq.~(\ref{Total_F}), one readily acquires the paramagnetic
free energy $f_{para}=-\frac{1}{\beta}(\log
2+\frac{1}{R}\log\cosh\beta)$ and the entropy
$s_{para}=\frac{1}{R}(\log\cosh\beta-\beta\tanh\beta)+\log 2$. In
zero temperature limit, $s_{para}=(1-\frac{1}{R})\log 2$. Since the
entropy becomes negative when $R<1$, the paramagnetic solution is
irrelevant for the error-correcting purpose.  As to the spin glass
phase, therefore, the replica symmetry should be broken, and a
simple assumption (frozen spins assumption) is adopted to avoid the
negative entropy\cite{Kaba01,Kaba03}, i.e.,  for low enough
temperature, the system settles in a completely frozen glassy phase.
On the transition boundary, both the frozen glassy phase and
paramagnetic phase share the identical free energy, and the
transition temperature is determined by $s_{para}(\beta_{g})=0$.
When $T<T_{g}$, the spin glass phase takes over, and the
corresponding free energy density can be written as
$f_{sg}=f_{para}(\beta_{g})$, independent of the temperature.
Besides the paramagnetic solution, there exists a ferromagnetic
solution ($m=1$). This solution is possible only in the case of
$K\rightarrow \infty$ (note that $R$ is kept finite). The related
ferromagnetic free energy with vanishing entropy could be derived
according to Eq.~(\ref{Total_F}), i.e.,
$f_{ferro}=-\frac{1}{R}(1-2p)$, independent of the temperature as
well. By identifying $f_{ferro}$ with $f_{sg}$, one can recover the
Shannon's bound as predicted by Shannon's channel encoding theorem,
implying $p_{c}\simeq 0.110028$ when $R=0.5$ (c.f.
Fig.~\ref{Regular}(a), the arrow indicates this critical noise
level). We report the phase diagram for the $K\rightarrow \infty$
code in Fig.~\ref{PDK}, note that the code rate is still kept to be
finite. It is important to remark that when the finite connectivity
is considered, modest loss in the final decoding quality should be
paid, i.e., the decoding overlap will be smaller than unity. To
illustrate the phase transition in the finite connectivity case, we
refer to the phase with finite decoding overlap as the ferromagnetic
phase. The transition is determined by identifying the ferromagnetic
free energy with 1RSB frozen spins free energy, then the glassy
phase ($m=0$) sets in to replace the ferromagnetic phase (finite
$m$). The corresponding critical noise level is obviously smaller
than the point where the magnetization (more precisely the decoding
overlap) drops to zero.

To solve Eq.~(\ref{BP}), population dynamics technique introduced in
Ref.~\cite{Cavity} is applied. The size of population is taken to be
of order $10^{4}$. Results are reported in Fig.~\ref{Regular}. For
$K=2$, no prior knowledge of the original message is required for
decoding, and the phase transition is of second order. As shown in
Fig.~\ref{Regular}(a), the critical noise level is determined by the
point where the 1RSB frozen spins free energy coincides with the RS
free energy. After the transition, the spin glass phase dominates
and the corresponding free energy is fixed to be $f_{sg}$.
Conversely, the phase transition is of first order for $K=3$, and
there is a remarkable drop in the free energy profile. However, the
computed free energy, which seems to be lower than the frozen spins
one, is unphysical after the phase transition because of its
corresponding negative entropy. Therefore the RS assumption is
incorrect and many states assumption should be adopted. The
performance of finite temperature decoding is also shown in
Fig.~\ref{Regular}(b). The decoding temperature is chosen to be the
optimal one, $\beta_{p}=\frac{1}{2}\log\frac{1-p}{p}$ named
Nishimori temperature \cite{Nishi_book,Rujan}. In this case, the
thermal temperature is identical to the noise temperature, and it is
observed that the performance is better than that of
zero-temperature decoding. Actually, the average spin alignment $m$
of decoding at Nishimori temperature sets an upper bound for all
achievable alignments \cite{Rujan}. As our numerical simulation has
shown, only the Nishimori temperature survives to get high overlap
when the critical noise level is approached. In contrast to the
$K=2$ case, the case of $K=3$ improves the decoding performance
significantly. However, the basin of attraction (BOA) shrinks
dramatically. We have to assume initial bias $m_{I}=0.8$ for finite
temperature decoding and $m_{I}=0.75$ for zero temperature decoding.
The compromise between good dynamical properties on one side ($K=2$)
and good performance on the other side ($K=3$) triggered us to
investigate the properties of the combined system with various $K$.

To further improve the decoding performance in the limit when the
temperature goes to zero, we have proposed the ECFP equation in
Sec.~\ref{subsec:ECFP}. The decoding overlap is plotted against the
flip rate in Fig.~\ref{ECFP}. Results obtained by WP are also shown
for comparison. Apparently, the decoding performance is improved
within an intermediate range of flip rate. In the presence of weak
noise, most of the propagating cavity fields take values larger than
$2$ and the energetic contribution plays a dominant role. Thus both
methods lead to identical performance. Once the noise becomes no
more small, the decoding performance achieved by ECFP starts to
deteriorate due to the divergence of some of the evanescent fields.
If we set a cutoff (e.g., $4.0$), to our surprise, the problem
mentioned above can be successfully circumvented. As shown in
Fig.~\ref{ECFP}, the result indeed outperforms that obtained by WP
which neglects the entropic effects. This can be understood as
follows, as flip rate becomes high enough, the relevant cavity
fields with $|I_{i\rightarrow a}|=1$ or $I_{i\rightarrow a}=0$,
emerge and contribute to the entropic effects \cite{Entropy}. These
information, omitted by WP, is correctly extracted by ECFP
procedure, and the decoding performance is finally boosted.
According to our numerical simulations, the value (e.g.,
$3.0,4.0,5.0$) we choose for the cutoff doesnot affect the decoding
results. When zero temperature decoding is concerned, we have
observed that ECFP is able to do a better job than WP since the
entropic effects have been incorporated. However, its decoding
performance still lies beneath that achieved by the optimal decoding
(MPM) where the decoding temperature is chosen to be the Nishimori
type. However, for MPM, one has to have a prior knowledge of the
channel noise, i.e., the flip rate of the noise. We present the
comparison between these two different kinds of decoding in
Fig.~\ref{MPMvsECFP}. In order to validate the mean field results,
we run the ECFP decoding algorithm on a single instance. The
comparison is shown in Fig.~\ref{ECFP}. The size of the graph is set
to be $N=10000$ and the code rate $R=0.5$. For one iteration step,
messages from each bit on the graph are updated one time on average.
We also set the maximal number of iteration steps $\mathcal{T}$ to
be $500$. The decoding result on single graph is averaged over ten
individual simulations for each flip rate $p$. As observed in our
simulations, the number of iteration steps, around $p=0.12$, exceeds
the preset value on most of the presented instances, which manifests
the ECFP starts to lose the convergence on a single graph. However,
the agreement with the mean field result is indeed remarkable.

\begin{center}
\begin{figure}
\epsfig{file=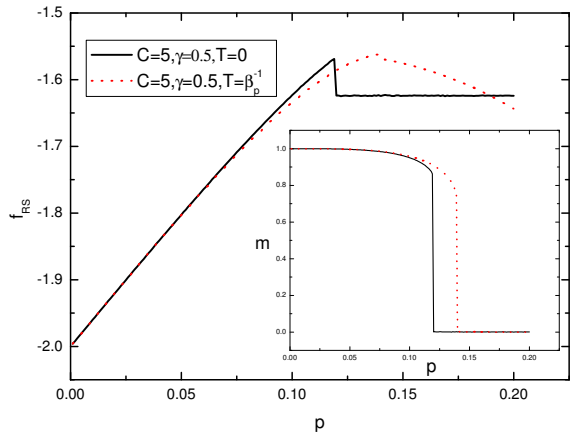,bb=10 13 279 214,width=8.0cm}
\caption{The decoding performance for irregular Sourlas codes with
$R=0.5$. The calculated mean values are shown and the corresponding
variances are smaller than the symbol size. The solid line
corresponds to zero temperature decoding while the dotted line
finite temperature decoding. Inset: the overlap versus flip rate.}
\label{Irregular}
\end{figure}
\end{center}

\subsection{Irregular Sourlas codes}

As defined above, the irregular Sourlas code is a combined system
with various values of $K$. It takes well the trade-off between
excellent convergence property of low-$K$ codes and high decoding
performance of high-$K$ codes. From the algorithmic point of view,
the irregular code is also termed cascading code put forward in
Ref.~\cite{KS} and further studied in Ref.~\cite{Hat}. In this
section, we report results on typical properties of the combined
system based on the cavity analysis presented in
Sec.~\ref{subsec:DISC}.

To retain the same code rate $R=0.5$, we choose $C=5, \gamma=0.5$ as
code construction parameters. Population dynamics recipe is used to
solve Eq.~(\ref{BP_Irre}), and the size of population is assumed to
be of order $10^{4}$. As shown in Fig.~\ref{Irregular}, the combined
system exhibits a first order phase transition as the $K=3$ case,
which was also observed in Ref. \cite{Hat} where the cascaded
encoding/decoding scheme was employed. After this transition, the
free energy crosses over to a lower value. However, as the flip rate $p$
increases to a high enough value, the RS entropy will be negative and the RS assumption is then incorrect,
indicating replica symmetry should be broken. As observed in
Fig.~\ref{Irregular}, the finite temperature (Nishimori's
temperature) decoding is superior to the zero temperature one when
the noise level becomes no longer low. Compared with the regular
code of $K=3$, the BOA for the combined system becomes larger thus
we only need to take the initial bias $m_{I}=0.6$. Additionally, the
overlap of decoding for the combined system is higher than that of
$K=2$. Therefore, results demonstrated in Fig.~\ref{Irregular}
provide us an opportunity to construct an optimal code. As has been
stated in Refs.~\cite{Hat,KS}, one can use multiple values of $K$ in
the interactions. As a first step, belief propagation or 1RSB
algorithm is run on a partial system with only low $K$ ($K=2$)
interactions since the low-$K$ code has perfect convergence
properties. The end overlap at the first stage is expected to be
well within the BOA of the combined system. Once higher body (e.g.,
$K=3$) interactions are invoked, an end overlap higher than the one
obtained by the initial step will be resulted in.

\section{Conclusions and Future Perspectives}
\label{sec_Con}

In this work, we have studied the finite connectivity Sourlas code
based on the cavity method. Conventional replica results on the
regular code are cross-checked. Moreover, this cavity analysis is
extended to the irregular case. Typical properties of the combined
system are investigated. It is shown that the decoding for the
combined system exhibits a first order phase transition as occurs in
the regular case ($K=3$). The combined system is of two striking
features, one is the initial bias required for convergence is
degraded, the other is the final performance is enhanced. Actually,
this does mean that the good dynamical properties (large BOA) and
high decoding performance should be compromised in the algorithmic
implementation. Thus introducing gradually higher $K$ interactions
seems to be an effective way to take advantage of this trade-off.

As for the regular codes system, the evanescent cavity fields
propagation equation is proposed for the first time. And it is
capable of extracting the entropic information in the zero
temperature limit, thus the decoding performance is considerably
enhanced compared with the traditional case where only the hard
field is taken into account. Numerical simulations on single
instances are compatible with the mean field results.

The cavity methodology, applied in our work, is very promising.
Unlike replica trick, it formulates assumptions in a more explicit
manner, even opens the way to algorithmic implementations on one
single instance. In this work, we also discovered that the system
shows negative entropy in the presence of low enough decoding
temperature and high enough flip rate, therefore 1RSB is needed for
further investigation on the finite connectivity Sourlas code.
Fortunately, the cavity method can be easily generalized to 1RSB
case. Meanwhile, the 1RSB frozen spin glass scheme we have adopted
in Sec.~\ref{sec_RSC} could be also cross-checked. On the other
hand, further study is required for the combined system to elucidate
under what conditions the channel capacity is achieved
\cite{Cap_KS}. Finally, the methodology is expected to be applied to
more practical codes like LDPC codes. These lines of research are
currently under way and these further investigations are anticipated
to provide deeper insights into a variety of codes with low density
nature of constructions.

\section*{Acknowledgments}

We thank Pan Zhang and Jie Zhou for stimulating discussions, and we
are grateful to anonymous referees for many helpful comments. The
present work was in part supported by the National Science
Foundation of China (Grant No. 10774150) and by the National Basic
Research Program (973-Program) of China (Grant No. 2007CB935903).

\end{document}